\documentclass{PoS}

\newcommand\onelineequation[2]{%
\begin{equation}\label{#1}
#2
\end{equation}%
}
\newcommand\twolineequation[4]{%
\begin{eqnarray}
#2
\label{#1}\\
#4
\label{#3}
\end{eqnarray}%
}
\newcommand\onefigure[4]{%
\begin{figure}[#4]
\centering
\includegraphics[width=#2]{#1}
\caption{#3}\label{#1}
\end{figure}%
}

\title{Testing the Yang--Mills vacuum wave functional Ansatz in 3$\,$+$\,$1 dimensions%
\addtocounter{footnote}{1}\thanks{This research was supported in part by the U.S.\ DOE under Grant No.\ DE-FG03-92ER40711 (J.G.), by the Slovak Research and Development Agency under
Contract No.\ APVV--0050--11, by the Slovak Grant Agency for Science, Projects VEGA No.\ 2/0070/09 and 2/0072/13, and by ERDF OP R\&D, Project meta-QUTE ITMS 26240120022 (\v{S}.O.).}}

\ShortTitle{Testing the Yang--Mills vacuum wave functional Ansatz in 3\,+\,1 dimensions} 

\author{Jeff Greensite\\
Niels Bohr International Academy, Blegdamsvej 17, DK--2100 Copenhagen \O, Denmark\\ and\\
Physics and Astronomy Dept., San Francisco State University, San Francisco, CA 94132, USA\\
E-mail: \email{jgreensite@gmail.com}}

\author{\addtocounter{footnote}{-2}
				\speaker{\v{S}tefan Olejn{\'\i}k}\\
        Institute of Physics, Slovak Academy of Sciences, SK--845 11 Bratislava, Slovakia\\
        E-mail: \email{stefan.olejnik@savba.sk}} 

\abstract{We proposed a simple Ansatz for the vacuum wave functional (VWF) of SU(2) gauge theory in temporal gauge. 
In 2\,+\,1 dimensions, the Ansatz was shown to be a fairly good approximation to the true VWF of the theory. Relative 
probabilities of various test configurations in the vacuum can be computed in numerical simulations of lattice-regulated SU(2) 
gauge theory by the method proposed long ago by Greensite and Iwasaki. We report promising (albeit still preliminary) results of testing the proposed VWF  in 3\,+\,1 dimensions on various sets of lattice gauge field configurations.}

\FullConference{Xth Quark Confinement and the Hadron Spectrum\\
		 8--12 October 2012\\
		 TUM Campus Garching, Munich, Germany}

\begin{document}

\section{Introduction: An Ansatz for the vacuum wave functional of SU(2) Yang--Mills theory in temporal gauge}
	The problem addressed in the present contribution is quite old, having been studied since at least 1979 \cite{Greensite:1979yn}. It can be stated very simply: In the Hamiltonian formulation of SU(2) Yang--Mills (YM) theory in temporal gauge in $D=d+1$ dimensions, one tries to find the ground-state (vacuum) wave functional (VWF) satisfying the YM Schr\"odinger equation plus the Gau\ss\ law constraint:
\twolineequation%
{1a}{\hat{\cal H}\Psi_0[A]=\int d^d x\left\lbrace-\frac{1}{2}\frac{\delta^2}{\delta A_k^a(x)^2}+\frac{1}{4}F_{ij}^a(x)^2\right\rbrace\Psi_0[A]&=&E_0\Psi_0[A],}
{1b}{\left(\delta^{ac}\partial_k+g\epsilon^{abc}A_k^b\right)\frac{\delta}{\delta A_k^c}\Psi_0[A]&=&0,}
hoping to be able to extract from the VWF useful information on color confinement and relevance of various gauge-field configurations in its mechanism. As simple as it appears, the problem defies satisfactory solution, despite attempts using various techniques\footnote{Some approaches to the problem are reviewed in \cite{Greensite:2011pj}, references to original papers can also be found therein.} (to appreciate the complications involved, see e.g.\ the horrifying expressions one deals with when looking for a solution perturbatively \cite{Krug:2012aa}).

	Some time ago, we proposed a simple Ansatz for an approximate VWF in temporal gauge \cite{Greensite:2007ij}:
\onelineequation{2}{\Psi_0[A]{=}{\cal{N}}
\exp\left[-\frac{1}{2}\displaystyle\int d^dx\,d^dy\, 
F^a_{ij}(x){\displaystyle
\left(\frac{1}{\sqrt{-{\cal D}^2-
\lambda_0+m^2}}\right)_{xy}^{ab}}F^b_{ij}(y)\right].}
In this expression, $F^a_{ij}(x)$ are components of the chromomagnetic field strength, ${\cal{D}}^2$ denotes the covariant laplacian in the adjoint representation, $\lambda_0$ is its lowest eigenvalue, and $m$ is a parameter with the dimension of mass. 

	In $D=2\,+\,1$ spacetime dimensions, the above VWF is particularly simple. There exists a single magnetic field with 3 color components, $B^a(x)\equiv F^a_{12}(x)$, $a=1,2,3$, the square of the coupling constant $g^2$ has the dimension of mass, and the parameter $m$ is to be chosen proportional to $g^2$. We have shown that the proposed VWF Ansatz is a fairly good approximation to the true ground state of the theory. Analytical and numerical evidence in its favor is briefly summarized in Sect.\ \ref{section2}.

	We conjecture that the expression (\ref{2}) could be a reasonable approximation to the true VWF also in $D=3\,+\,1$ dimensions. However, to justify this proposition is much more complicated. Fortunately, there exists a method of measuring the (square of the) VWF in numerical simulations, proposed long ago by Greensite and Iwasaki \cite{Greensite:1988rr}, and utilized recently also for testing the wave functional in $D=2\,+\,1$ dimensions \cite{Greensite:2011pj}. The method is outlined in Sect.\ \ref{section3} and the first preliminary results for test configurations  in $D=3\,+\,1$ dimensions are presented in Sect.\ \ref{section4}. They are promising, but conclusions (Sect.\ \ref{section5}) have still to be taken with some caution.  

\section{Summary of results for (2$\,$+$\,$1) dimensions}\label{section2}
	By construction, the VWF of Eq.\ (\ref{2}) in $(2\,+\,1)$ dimensions reproduces the well-known solution of the Schr\"odinger equation for the ground state of electrodynamics in the free-field limit ($g\to 0$). In addition, we have shown that it has a number of other attractive features:

	1.\ It is a good approximation to the true vacuum also for strong fields constant in space and varying only in time \cite{Greensite:2007ij}.

	2.\ For slowly varying chromomagnetic fields it reduces to the ``dimensional-reduction'' form:
\onelineequation{3}{\Psi_0\sim\exp\left[-\textstyle{\frac{\mu}{2}}{\displaystyle\int} d^2x\,B^a_\mathrm{slow}(x)B^a_\mathrm{slow}(x)\right]}
that has been argued to be the correct form of the Yang--Mills VWF at large scales \cite{Greensite:1979yn,Halpern:1978ik}.

	3.\ It exhibits confinement (non-zero string tension) for the parameter $m\ne 0$, and the non-zero value of $m$ seems preferred in a variational estimation of the vacuum energy \cite{Greensite:2007ij}.

	4.\ We proposed a recursion method of generating field configurations distributed according to the proposed VWF \cite{Greensite:2007ij} and compared a few quantities computed in an ensemble of recursion configurations with those obtained in ensembles of true SU(2) gauge-field configurations. The values of the mass gap \cite{Greensite:2007ij}, the ghost propagator in Coulomb gauge, and the color-Coulomb potential~\cite{Greensite:2010tm} in recursion configurations reasonably agreed with true configurations.

	5.\ We computed the square of the true VWF for sets of simple test configurations (by the method described in some detail in Sect.\ \ref{section3}),  the results were consistent with expectations based on the proposed VWF, Eq.\ (\ref{2}) \cite{Greensite:2011pj}.

	The above results led us to conclude that the proposed VWF, in $(2\,+\,1)$ dimensions, agrees with the true Yang--Mills vacuum wave functional for the bulk of the probability distribution, with possibly a small disagreement in the tail of the distribution.

	However, we live in $D=3\,+\,1$ spacetime dimensions, and the success of our wave functional (\ref{2}) in the real world is by far not guaranteed. If we postulate the same form to work also in $D=3\,+\,1$ dimensions, there are some good and some bad news. The VWF again reduces to the QED VWF in the $g\to 0$ limit (assuming $m$ can be neglected as well), and also points 1 and 2 above still remain valid \cite{Greensite:2007ij}. Unfortunately, the recursion method invented for generating gauge configurations in $D=2\,+\,1$ does not work in $D=3\,+\,1$. It profited from the possibility of easily going from potentials~$A_k$ to field strengths $B$ and back (in a variant of axial gauge), which is not possible in $D=3\,+\,1$ because of Bianchi constraints. We are forced (at least for the moment) to test the proposed VWF only in numerical simulations on some sets of test configurations.

\section{Direct measurement of the VWF in numerical simulations}\label{section3}

	There exists a method of direct measurement of the square of the VWF for some sets of gauge-field configurations in numerical simulations of the $D=4$ Euclidean pure Yang--Mills theory. The method was invented by Greensite and Iwasaki \cite{Greensite:1988rr}, and is based on the observation that the square of the VWF for a certain gauge-field configuration $U'(x)$ at fixed time (say $t=0$) is expressed through a path integral:
\onelineequation{4}{\vert\Psi[U']\vert^2=\displaystyle\frac{1}{Z}\int [DU]\delta(U_0)\prod_x \delta\left[U(x,0)-U'(x)\right]e^{-S}.}
Now take a set of configurations ${\cal{U}}=\left\{U^{(j)}(x), j=1,2,\dots,M\right\}$. The value of each
$\left\vert\Psi_0\left[U^{(j)}\right]\right\vert^2$ is proportional to $P_j$, the probability that a lattice configuration on the $t=0$ time slice is equal to the $j$-th configuration $U^{(j)}(x)\in\cal{U}$. That probability can be computed numerically by a modified Monte Carlo simulation. All links at $t\ne0$ are updated in the usual way, e.g.\ by heat bath for the SU(2) Wilson action. At $t=0$, one configuration from the set $\cal{U}$ is chosen randomly, and accepted/rejected by the usual Metropolis algorithm. Then
\onelineequation{5}{\vert\Psi[U^{(j)}]\vert^2\propto\lim_{N_\mathrm{tot}\to\infty}\,\frac{N_j}{N_\mathrm{tot}}\,,}
where $N_j$ is the number of times that -- in a given simulation -- the $j$-th configuration in the set is selected by Metropolis,
and $N_\mathrm{tot}$ denotes the total number of updates of the $t=0$ plane.

	A limitation of the present method is that $\left\vert\Psi_0\left[U^{(j)}\right]\right\vert^2$ falls off exponentially with the action of the configuration, so to achieve reasonable acceptance rate in Metropolis accept/reject steps all configurations in a test set have to be relatively close in action. (This is regulated by parameters $\kappa$, $\alpha$ and $\gamma$ in sets of configurations below.)

\section{Preliminary results for test configurations in (3\ +\ 1) dimensions}\label{section4}
	In this pilot study. we address a simple question in lattice simulations: Can the data distinguish between the dimensional-reduction form of the squared VWF%
\footnote{On a lattice, the integral in the exponent is proportional to $S_\mathrm{sp}=\sum_P\left(1-{\textstyle\frac{1}{2}}\mbox{Tr}[U_P]\right)$, where the sum runs over all spatial plaquettes.}
\onelineequation{6}{\vert\Psi_0\vert^2\;\sim\;\exp\left[-\mu{\displaystyle\int} d^3x\;B^a_{k}(x)B^a_{k}(x)\right]}
and that following from (\ref{2})
\onelineequation{7}{\vert\Psi_0\vert^2\;\sim\;\exp\left[-\displaystyle\int d^3x\,d^3y\;B^a_{k}(x){\displaystyle\left(\frac{1}{\sqrt{-{\cal D}^2-
\lambda_0+m^2}}\right)_{xy}^{ab}}B^b_{k}(y)\right]\;?}
What fits better to the true vacuum of the pure YM theory?\footnote{The true VWF cannot just be of the dimensional-reduction form for all field configurations. This would lead to exact, not just approximate, Casimir scaling, and incorrect results at short distances/high frequencies \cite{Greensite:2011pj,Quandt:2010yq}.}

	To investigate this question, we computed, by the method described in Sect.\ \ref{section3}, the squared-VWF values for three types of configurations:

	1.\ non-abelian constant configurations;

	2.\ abelian plane waves on lattices of variable size with maximum wave-length ($\lambda=L$);

	3.\ abelian plane waves on a fixed-size lattice with variable wave-length.\\
Simulations were performed for coupling constants $\beta=2.2, 2.3, 2.4,$ and $2.5$ and lattices of sizes $L=12, 16, 20, 24,$ and $28$. We present only a subset of data; more results with higher statistics and systematic error analysis will be presented elsewhere \cite{Greensite:2013zz}.

\paragraph{\textit{Non-abelian constant configurations:}} We used configurations
\onelineequation{8}{{\cal{U}}_\mathrm{NAC}=\left\{U_k^{(j)}(x)=\sqrt{1-(a^{(j)})^2}\;\mathbf{1}_2+i a^{(j)}\;\sigma_k\right\}\quad\mbox{with}\quad
a^{(j)}=\displaystyle\sqrt[4]{\frac{\kappa}{6L^3}j\ },}
for which $S_\mathrm{sp}=\kappa j$. These are not useful to address the above question, because of the (approximate) equality, valid for constant configurations:
\onelineequation{9}{B^a_k(x)\left(\frac{1}{\sqrt{-{\cal{D}}[A]^2-\lambda_0+m^2}}\right)_{xy}^{ab}B^b_k(y)\approx
\frac{2}{m}S_\mathrm{sp}\,.}
(Sum over repeated indices is implied.) We used therefore these sets of configurations only to ``calibrate'' our computer code, i.e.\ for comparison with earlier data of Greensite and Iwasaki, which used modest-size lattices, $8^4$ and $10^4$. New data, from $16^4$ and $20^4$ lattices, are fully compatible with the old ones. 

\paragraph{\textit{Abelian plane waves:}} The simplest choice is
\onelineequation{10}{{\cal{U}}_\mathrm{APW}=\left\{ U_1^{(j)}(x)=\sqrt{1-\left(a_{n}^{(j)}(x)\right)^2}\;\mathbf{1}_2+i a_{n}^{(j)}(x)\;\sigma_3;\qquad U_2^{(j)}(x)=U_3^{(j)}(x)=\mathbf{1}_2\right\}}
with
\onelineequation{11}{a^{(j)}(x)=\displaystyle\sqrt{\frac{\alpha_{{n}}+\gamma_{{n}} j}{L^3}}\cos\left(\frac{2\pi x_2}{\lambda_{{n}}}\right)\qquad\mbox{and}\qquad\lambda_{{n}}=\frac{L}{{n}}\,.}
For such configurations one expects
\onelineequation{12}{-\log\left(\displaystyle\frac{N_j}{N_\mathrm{tot}}\right)\propto {\textstyle\frac{1}{2}}(\alpha_n+\gamma_n j)\times\omega(\lambda_n),}
where
\onelineequation{13}{\omega(\lambda_n)=
\left\{\begin{array}{c c l}
\frac{1}{2} \mu k^2(\lambda_n) & \;\dots & \mbox{from DR $\vert\Psi_0\vert^2$, Eq.\ (\protect\ref{6}),}\\[3mm]
\displaystyle\frac{k^2(\lambda_n)}{\sqrt{k^2(\lambda_n)+m^2}}& \;\dots & \mbox{from our $\vert\Psi_0\vert^2$, Eq.\ (\protect\ref{7}),} 
\end{array}\right.\quad\mbox{and}\quad k^2(\lambda_n)\equiv2\left(1-\cos\frac{2\pi}{\lambda_n}\right).}

	At a fixed coupling $\beta$, fixed lattice size $L$ and a certain wavelength of $\lambda_n$, $-\log\left({N_j}/{N_\mathrm{tot}}\right)$ is a linear function
of $j$, and one can get $\omega(\lambda_n)$ from the best fit of the form (\ref{12}). In principle the obtained slope $\omega$ could depend on the choice of constants $\alpha_n$ and $\gamma_n$ that parametrize the set of configurations. It turns out that slopes coming from quite different pairs $(\alpha_n, \gamma_n)$ are compatible within errors. This is illustrated in Figure \ref{2-parameter-sets}. One should however keep in mind that the ambiguity of the choice of $(\alpha_n, \gamma_n)$ is a source of systematic error that has not yet been accounted for in the analysis of data presented below.

\onefigure{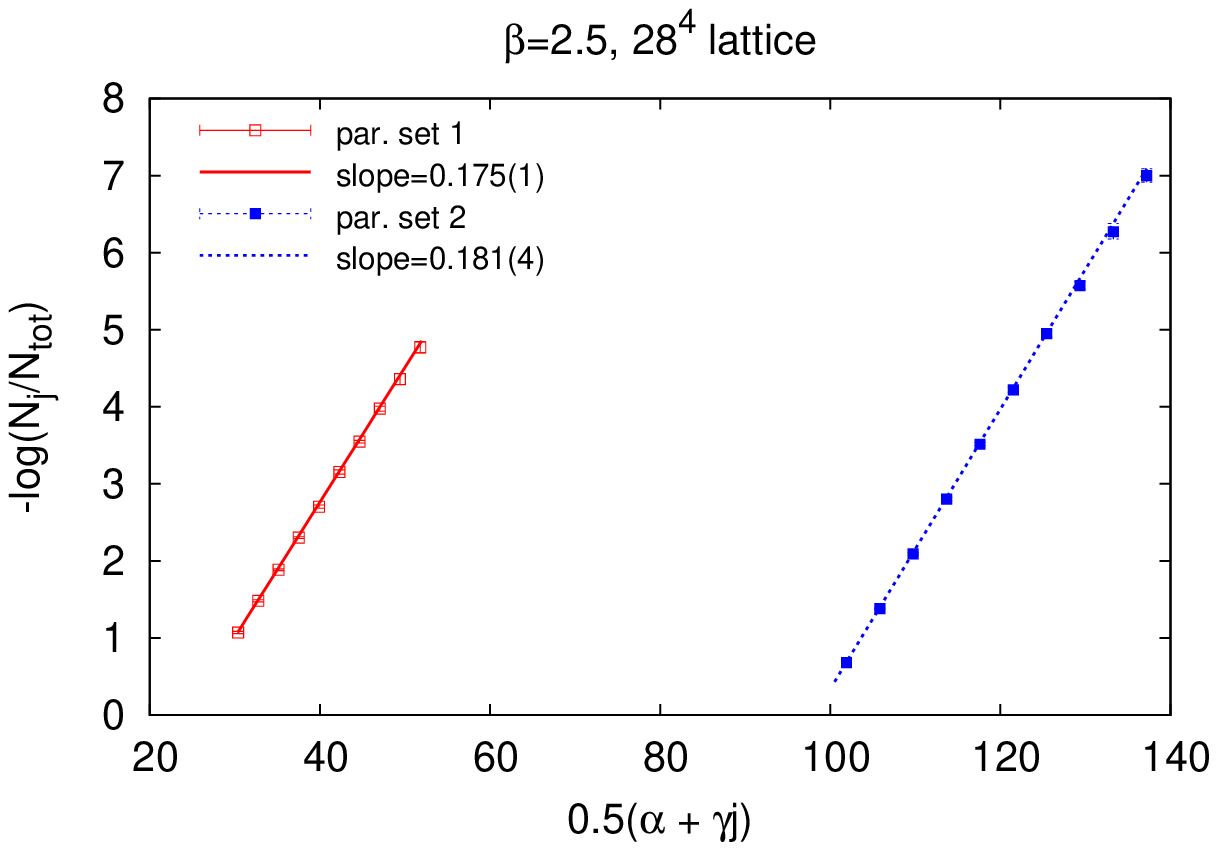}{8.2cm}{The slope of the linear dependence of $-\log\left({N_j}/{N_\mathrm{tot}}\right)$ vs.\ ${\textstyle\frac{1}{2}}(\alpha+\gamma j)$ depends slightly on the chosen parameter set $(\alpha,\gamma)$. The displayed data are for abelian plane waves with $\lambda=L$ at $\beta=2.5$ and $L=28$.}{t!}


	Figure \ref{omega_vs_k_2p4} represents the results for \textbf{\textit{abelian plane waves with maximum wavelength $\lambda=L$ for a number of lattices sizes $L$}} at a fixed gauge coupling, $\beta=2.4$. The solid (red) line represents a fit of the form $\left(c\times k^2(L)/\sqrt{k^2(L)+m^2}\right)$ motivated by our proposed VWF. The data are described satisfactorily by this fit. However, even the fit of the form $\left(a+b k^2(L)\right)$ (motivated by the dimensional-reduction VWF) does represent the data quite reasonably (the dotted blue line), though the $\chi^2$ of the latter fit is considerably worse than of the former one.

\onefigure{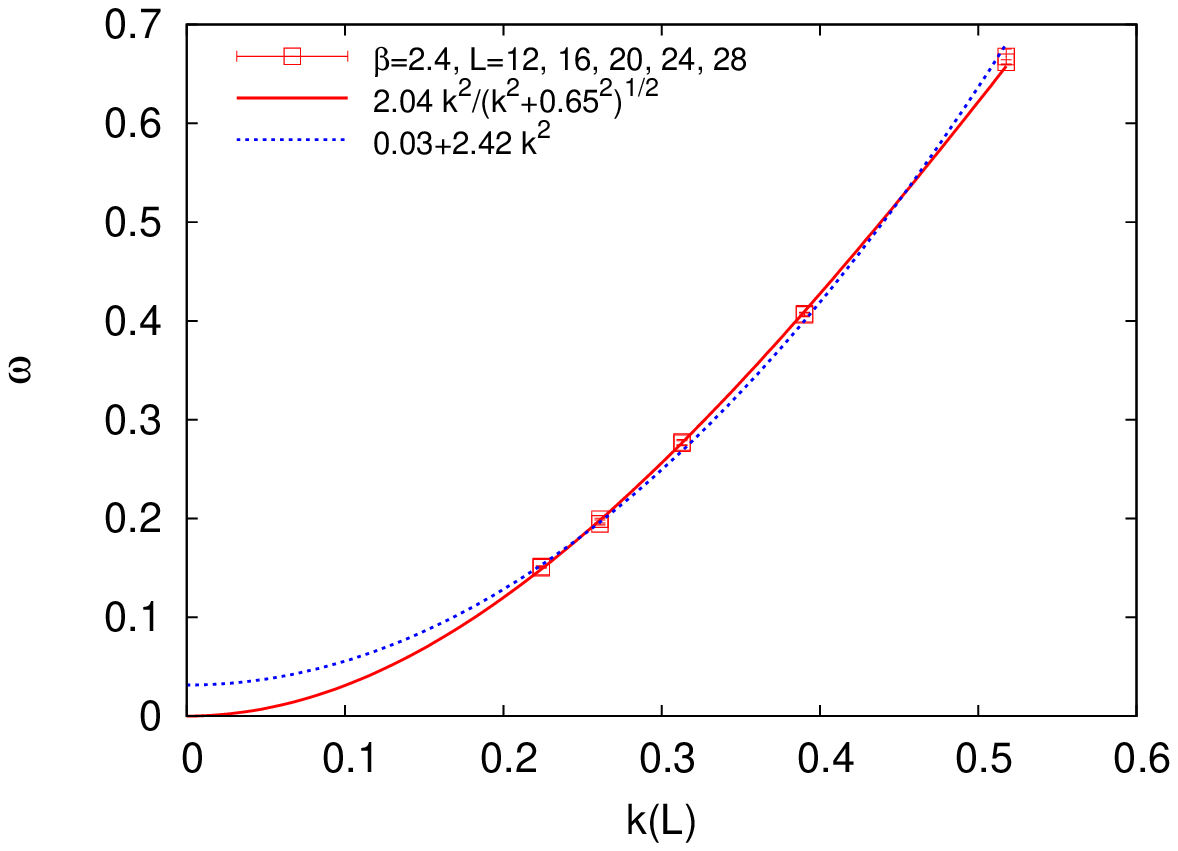}{8.2cm}{The quantity $\omega$ vs.\ $k(L)$ (see \protect\ref{13}) for abelian plane waves with maximum wave-length at $\beta=2.4$ for a number of lattice sizes $L$, together with fits inspired by the VWFs (\protect\ref{6}) and (\protect\ref{7}).}{b!}

	The parameter $m$ of the former fit, if it corresponds to a physical mass in the continuum limit, should scale properly as a function of the coupling $\beta$. The preliminary data look promising in this respect. If we divide $m$ by the value of another mass quantity, e.g.\ the square root of the string tension, at the same coupling, the ratio should not depend on the coupling. In Figure \ref{m_vs_beta_using_k2} we display such ratio at four values of gauge coupling in the scaling window, and the result is compatible with a constant, $m/\sqrt{\sigma}\approx2.36$. One should, however, take this result with some caution: the errors shown may be underestimated, a substantial change of $m$ can be compensated by a change of the fit parameter $c$ without deteriorating the agreement with data. This is illustrated in Figure \ref{omega_vs_k_2p2}. A fit to data with one parameter set $(\alpha, \gamma)$ at $\beta=2.2$  gave $m=1.45 (40)$, a different parameter
set $(\alpha, \gamma)$ leads to $m=1.07 (8)$, and the fit to combined data (that was displayed in Figure \ref{m_vs_beta_using_k2}) resulted in $m=1.17 (14)$. Still, the two fits with $m$  differing by $\sim$30\% are hardly distinguishable in Figure~\ref{omega_vs_k_2p2}!

\onefigure{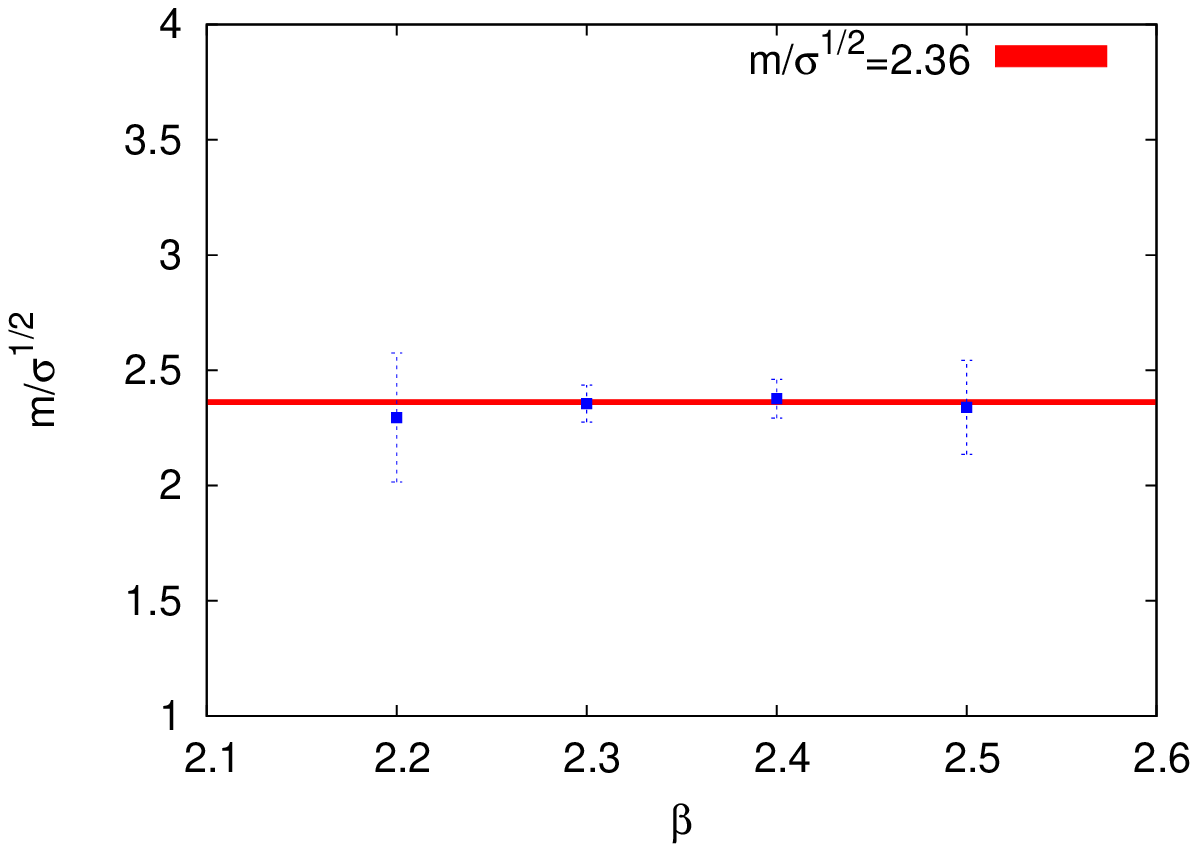}{8.2cm}{The ratio of the parameter $m$ (extracted from fits to data for abelian plane waves with maximum wave-length) to the square root of the string tension $\sigma$ vs. the coupling constant $\beta$. }{t!}

\onefigure{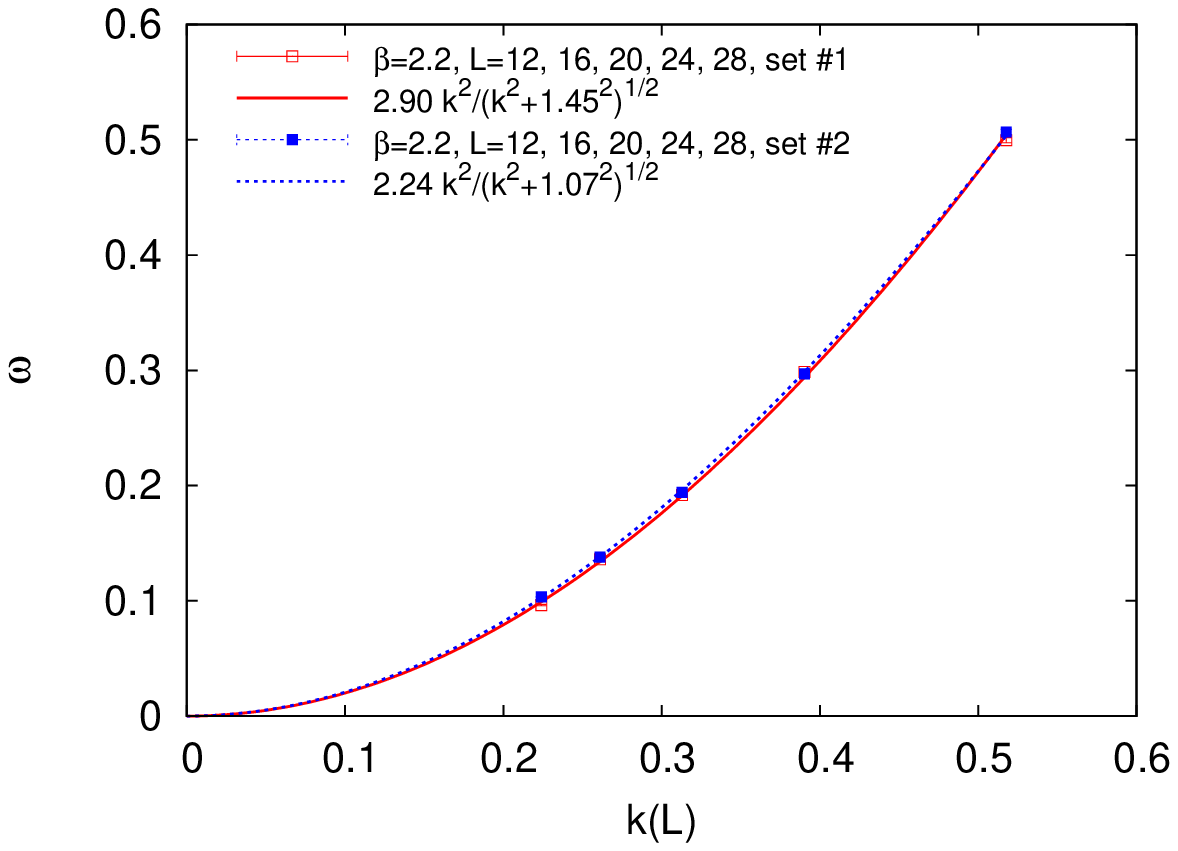}{8.2cm}{$\omega$ vs.\ $k(L)$ for abelian plane waves with maximum wave-length at $\beta=2.2$ for a number of lattice sizes $L$. Two fits to data sets corresponding to different choices of parameters $(\alpha,\gamma)$ give considerably different $m$ values.}{b!}

	To discriminate between the two forms in Eq.\ (\ref{13}) one needs to use sets of \textbf{\textit{abelian plane waves with variable wavelengths $\lambda_n$ on a fixed size lattice}}. An illustrative result is displayed in Figure \ref{omega_vs_k_2p4_l24}. Here the fit motivated by our proposed form of the VWF is clearly preferred.  However, the mass parameter $m$ resulting from the best fit to these preliminary data ($m\approx 1.29$) is somewhat different from the value obtained from abelian plane waves with maximum wavelength ($m\approx 0.90$). More extensive simulations are needed to resolve this discrepancy.


\onefigure{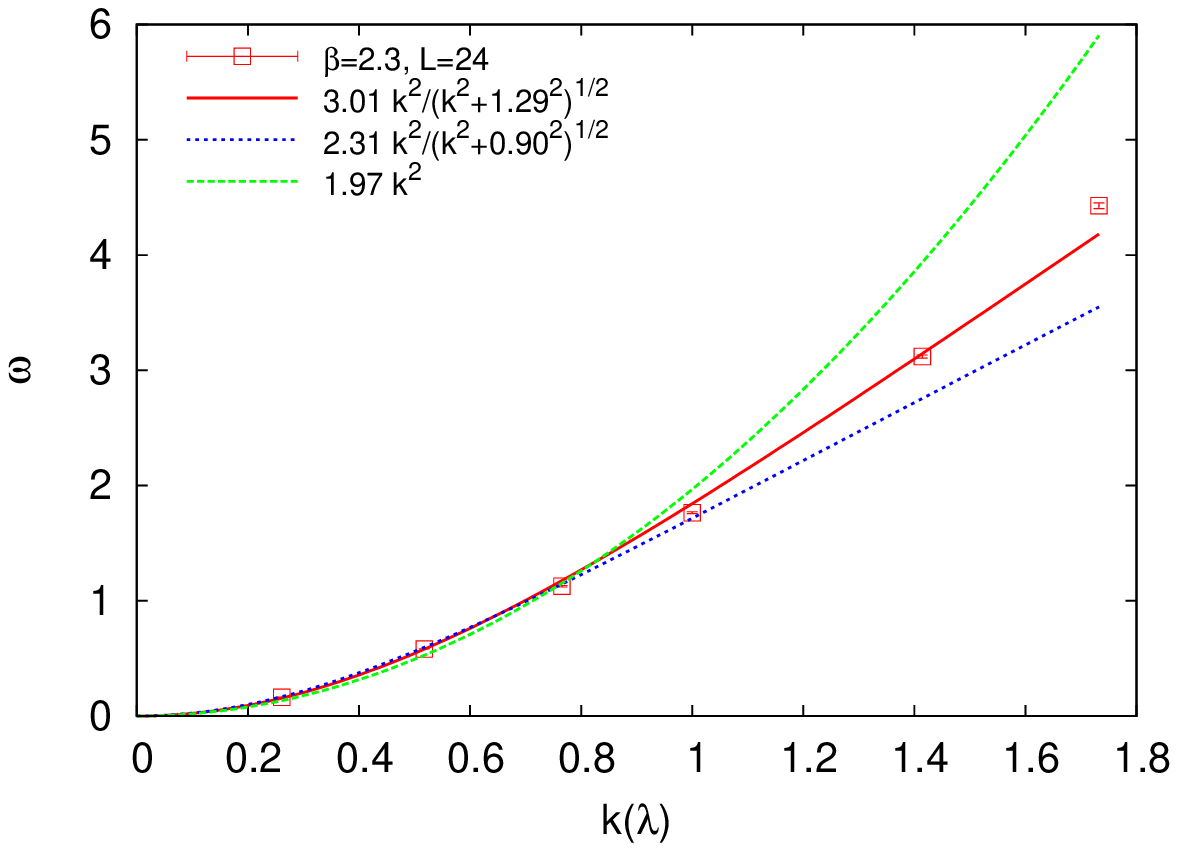}{8.2cm}{$\omega$ vs.\ $k(\lambda)$ for abelian plane waves with variable wave-length at $\beta=2.3$ on $24^4$ lattice.}{t!}

\section{Conclusions}\label{section5}

	The preliminary investigation of the SU(2) Yang--Mills vacuum wave functional in $D=3\,+\,1$ dimensions has provided promising outcome:

	1.\ The results of this pilot study are consistent with the old ones by Greensite and Iwasaki~\cite{Greensite:1988rr}, but extend their investigation to larger lattices, and direct tests of the approximate form of the vacuum wave functional (\ref{2}).

	2.\ For the three sets of simple configurations (non-abelian constant configurations, abelian plane waves with $\lambda=L$, abelian plane waves with variable wavelengths) the computed values of $\vert\Psi_0\vert^2$ are consistent with the form (\ref{2}), but the evidence cannot be considered conclusive.

	3.\ The parameter $m$ of the VWF, determined from fits to probabilities of plane-wave configurations with $\lambda=L$, scales as a physical quantity with dimension of mass.

	4.\ The numerical data for abelian plane waves deviate from expectations based on the di\-men\-sion\-al-reduction form of the VWF.

	For the future, more extensive simulations, higher statistics and systematic error analysis for simple test configurations are called for. The most important (and difficult) task is to find an efficient method in $(3\,+\,1)$ dimensions for generating gauge-field configurations distributed according to the VWF (\ref{2}).

\end{document}